\documentclass[twocolumn,tighten]{aastex63}
\pdfoutput=1 %for arXiv submission
\usepackage{amsmath,amstext}
\usepackage[T1]{fontenc}
\usepackage{apjfonts} 
\usepackage[figure,figure*]{hypcap}
\usepackage{url}

\shorttitle{X-raying the Terzan 5 Millisecond Pulsars}
\shortauthors{Bogdanov et al.}

\begin{document}

\title{A DEEP \textit{CHANDRA X-RAY OBSERVATORY} STUDY OF THE MILLISECOND PULSAR POPULATION \\  IN THE GLOBULAR CLUSTER TERZAN 5}

\correspondingauthor{Slavko Bogdanov}
\email{slavko@astro.columbia.edu}

\author[0000-0002-9870-2742]{Slavko Bogdanov}
\affiliation{Columbia Astrophysics Laboratory, Columbia University, 550 West 120th Street, New York, NY 10027, USA}

\author[0000-0003-2506-6041]{Arash Bahramian}
\affiliation{International Centre for Radio Astronomy Research-Curtin University, GPO Box U1987, Perth, WA 6845, Australia}

\author[0000-0003-3944-6109]{Craig O.~Heinke}
\affiliation{Department of Physics, University of Alberta, CCIS 4-183, Edmonton, AB, T6G 2E1, Canada}

\author[0000-0001-5799-9714]{Scott M.~Ransom}
\affiliation{National Radio Astronomy Observatory, 520 Edgemont Road, Charlottesville, VA 22903, USA}

\begin{abstract}
We present an analysis of 745.6\,ks of archival \textit{Chandra X-ray Observatory} Advanced CCD Imaging Spectrometer data accumulated between 2000 and 2016 of the millisecond pulsar (MSP) population in the rich Galactic globular cluster Terzan 5. Eight of the 37 MSPs with precise positions are found to have plausible X-ray source matches. Despite the deep exposure, the remaining MSPs are either marginally detected or have no obvious X-ray counterparts, which can be attributed to the typically soft thermal spectra of rotation-powered MSPs, which are strongly attenuated by the high intervening absorbing column ($\sim10^{22}$\,cm$^{-2}$) towards the cluster, and in some instances severe source crowding/blending. For the ``redback'' MSP binaries, PSRs J1748$-$2446P and J1748$-$2446ad, and the ``black widow'' binary PSRs J1748$-$2446O, we find clear evidence for large-amplitude X-ray variability at the orbital period consistent with an intrabinary shock origin. The third redback MSP in the cluster, PSR J1748$-$2446A, shows large amplitude variations in flux on time scales of years, possibility due to state transitions or intense flaring episodes from the secondary star.

\end{abstract}


\section{Introduction}
The extreme number densities of stars in the cores of globular clusters ($\sim10^{5-6}$ M$_{\odot}$ pc$^{-3}$) are highly conducive to close dynamical interactions, making them veritable factories of compact binaries \citep[e.g.,][]{Pooley03}.  These include quiescent low-mass X-ray binaries (qLMXBs; \citealt{Rutledge01}); cataclysmic variables (CVs; \citealt{Cool95}); chromospherically active main-sequence binaries (ABs; \citealt{Edmonds03}); and radio millisecond pulsars \citep[][]{Bogdanov06,Bogdanov10,Bogdanov11}. In recent years, an additional class of exotic globular cluster  binaries have been identified -- candidate black hole X-ray binaries \citep{Strader12,Chomiuk13,Miller-Jones15}. As these various classes of compact systems exhibit faint ($\le 10^{33}$ erg s$^{-1}$) X-ray emission, the \textit{Chandra X-ray Observatory}, owing to its sub-arcsecond imaging capabilities, has proven to be the ideal tool for the identification and study of a multitude of such exotic objects in globular cluster cores  \citep{Heinke05,Bogdanov10,Bogdanov11,Forestell14}. 

Terzan 5 is a dense globular cluster located $\sim5.9$ kpc away \citep{Valenti07} in the direction of the Galactic center.  It is remarkable in that it has the largest collision rate of any globular cluster in the Galaxy \citep{Verbunt87,Bahramian13}. Thus, it is not surprising that Terzan 5 hosts three transiently luminous LMXBs, EXO 1745$-$248 \citep{Makishima81}, IGR J17480$-$2446 \citep{Papitto11}, and Swift J174805.3$-$244637 \citep{Bahramian14}, plus at least 10 additional quiescent LMXB candidates \citep{Heinke06}. Terzan 5 also has the distinction of hosting the largest number of known MSPs, currently 37 in total (for reference, the second highest is 47 Tuc with 25 MSPs \citealt{Camilo00,Ridolfi16,Pan16}).  \citet{Ransom05} first identified a large population of radio MSPs in Terzan 5, with a harvest of 21 new MSPs using a deep targeted radio survey with the Green Bank Telescope. Subsequent radio pulsation search campaigns have extended the population further with 16 new discoveries \citep{Prager2017,Andersen2018,Cadelano18}, bringing the total up to 38. Perhaps the most notable of the Terzan 5 MSPs is PSR J1748$-$2446ad (Ter5 ad hereafter), the fastest among known pulsars, with a spin of 716\,Hz (see \citealt{Hessels06}). Along with PSRs J1748$-$2446A (Ter5 A) and J1748$-$2446P (Ter5 P), Ter5 ad also belongs to the so-called ``redback'' variety of eclipsing MSPs, which are a growing subset of binary MSPs bound to relatively massive ($\ge0.2$ M$_{\odot}$) non-degenerate stars.  Redbacks have recently garnered a great deal of interest since one of them, PSR J1824$-$2452I in the globular cluster M28, was seen to switch dramatically between rotation-powered (radio) and accretion-powered (X-ray) pulsations \citep{Papitto13}, thereby lending strong support to the long-suspected evolutionary connection between low-mass X-ray binaries and recycled pulsars. Archival \textit{Chandra} data of M28 showed that PSR J1824$-$2452I underwent a transition to a low-luminosity ($10^{33}$ erg s$^{-1}$) LMXB state in 2008 as well \citep{Linares14}. In addition, two redbacks in the field of the Galaxy, XSS J12270--4859 and PSR J1023+0038, have been seen to undergo similar transformations \citep{Patruno14,Bassa14}. 

In X-rays, most rotation-powered MSPs in globular clusters and the field of the Galaxy show soft spectra \citep[see, e.g.,][]{Bogdanov06}; sensitive X-ray observations with adequate time resolution of nearby MSPs have revealed broad pulsations \citep[see][and references therein]{Guillot2019}, interpreted as thermal radiation arising from small regions on the surface. In contrast, the three most energetic MSPs known exhibit remarkably narrow pulses with hard power-law spectra \citep[e.g.,][]{Gotthelf2017}, indicative of non-thermal radiation from relativistic particles accelerated in the pulsar magnetosphere. The third broad category of X-ray emission from radio MSPs is non-thermal radiation that is modulated at the binary period \citep{Bogdanov05,Huang2012}, observed exclusively from the so-called ``black widow'' and redback varieties of eclipsing binary MSPs. This can be interpreted as high-energy radiation due to a shock driven by the interaction of the pulsar wind with material from a close companion star \citep{Arons93,Romani16,Wadiasingh17}.

The first attempt to characterize the X-ray source population of Terzan 5 with \textit{Chandra} ACIS in 2000 was hampered by a luminous outburst from the transient EXO~1745$-$248 \citep{Heinke03}.  Subsequently, in 2003 a 39.3-ks \textit{Chandra} exposure, 50 sources were detected down to a limiting 1--6 keV X-ray luminosity $\sim 10^{31}$ erg s$^{-1}$ within the half-mass radius of the cluster \citep{Heinke06}.  Thirty three of these sources have $L_X>10^{32}$ erg s$^{-1}$, the largest number seen in any globular cluster, with more than twice as many X-ray binaries in this luminosity range as NGC 6440 and NGC 6266, the next richest X-ray clusters studied so far \citep{Pooley02,Pooley03}.  In the mean time, a number of new \textit{Chandra} exposures of Terzan 5 have been performed for the purposes of monitoring   the cooling of the transiently accreting neutron stars Swift J174805.3$-$244637 \citep{Degenaar2015} and IGR~J17480$-$2446 \citep{Ootes2019}. \citet{Bahramian15} used the same exposures of Terzan 5 as part of a broader study of the long-term variability of quiescent LMXBs in globular clusters.  These observations were also used to identify a transitional millisecond pulsar candidate \citep[CX1, see][]{Bahramian2018}. Collectively, the \textit{Chandra} observations of Terzan 5 not affected by bright transients constitute an ultra-deep (745.6 ks check) archival data set spanning 16 years (2000--2016), which offers significantly improved constraints on the properties of its various X-ray source populations including its sizable sample of rotation-powered MSPs. A new comprehensive catalog of 212 X-ray point sources detected within the half-mass radius of Terzan 5 based on these observations was compiled in \citet{Bahramian2020}.

In this paper, we present X-ray imaging, spectroscopic, and variability analyses of the sample of MSPs in Terzan 5 using this rich archival \textit{Chandra X-ray Observatory} data set. The work is organized as follows. In Section~\ref{sec:obs}, we describe the observations and data reduction procedure. Section~\ref{sec:imaging} focuses on imaging , source detection, and the identification of X-ray counterparts to the MSPs. In Section~\ref{sec:spectral} we investigate the spectral properties of the X-ray sources, while in Section~\ref{sec:variability} we examine variability. We offer conclusions in Section~\ref{sec:conclusions}.

\begin{deluxetable}{lccr}
\tablecolumns{4} 
\tablewidth{0pt}  
\tablecaption{Log of \textit{Chandra} ACIS imaging observations of Terzan 5\label{tab:Obs}}  
\tablehead{\colhead{Date} & \colhead{ObsID} & \colhead{Detector} & \colhead{Exposure} \\ \colhead{} & \colhead{} & \colhead{} & \colhead{(ks)}}
\startdata
2000 Jul 29	&	654*	&	ACIS-I	&	5.0	  \\
2000 Jul 24	&	655*	&	ACIS-I	&	42.2  \\
2003 Jul 13	&	3798	&	ACIS-S	&	39.3   \\
2009 Jul 15	&	10059	&	ACIS-S	&	36.3   \\
2010 Oct 24	&	11051*	&	ACIS-S	&	9.9	\\
2011 Nov 3	&	12454*	&	ACIS-S	&	9.8	\\
2011 Feb 7  &   13225   &   ACIS-S  &   29.7   \\
2011 Apr 29 &   13252   &   ACIS-S  &   39.5   \\
2011 Sep 5	&	13705	&	ACIS-S	&	13.9   \\
2012 May 13	&	13706	&	ACIS-S	&	46.5   \\
2012 Jul 30	&	13708*	&	ACIS-S	&	9.8	   \\
2011 Sep 8	&	14339	&	ACIS-S	&	34.1   \\
2012 Sep 17 &   14475   &   ACIS-S &    30.5   \\
2012 Oct 28 &   14476   &   ACIS-S &    28.6   \\
2013 Feb 5  &   14477   &   ACIS-S &    28.6   \\
2013 Jul 16 &   14478   &   ACIS-S &    28.6   \\
2014 Jul 15 &   14479   &   ACIS-S &    28.6   \\
2013 Feb 22	&	14625	&	ACIS-S	&	49.2   \\
2013 Feb 23	&	15615	&	ACIS-S	&	84.2   \\
2014 Jul 20	&	15750	&	ACIS-S	&	23.0   \\
2014 Jul 17	&	16638	&	ACIS-S	&	71.6   \\
2016 Jul 13	&	17779	&	ACIS-S	&	68.9   \\
2016 Jul 15	&	18881	&	ACIS-S	&	64.7   \\
\enddata
\tablenotetext{*}{Observations affected by transient outbursts not used in this analysis.}
\end{deluxetable}

\begin{figure*}
\centering
\includegraphics[width=0.88\textwidth]{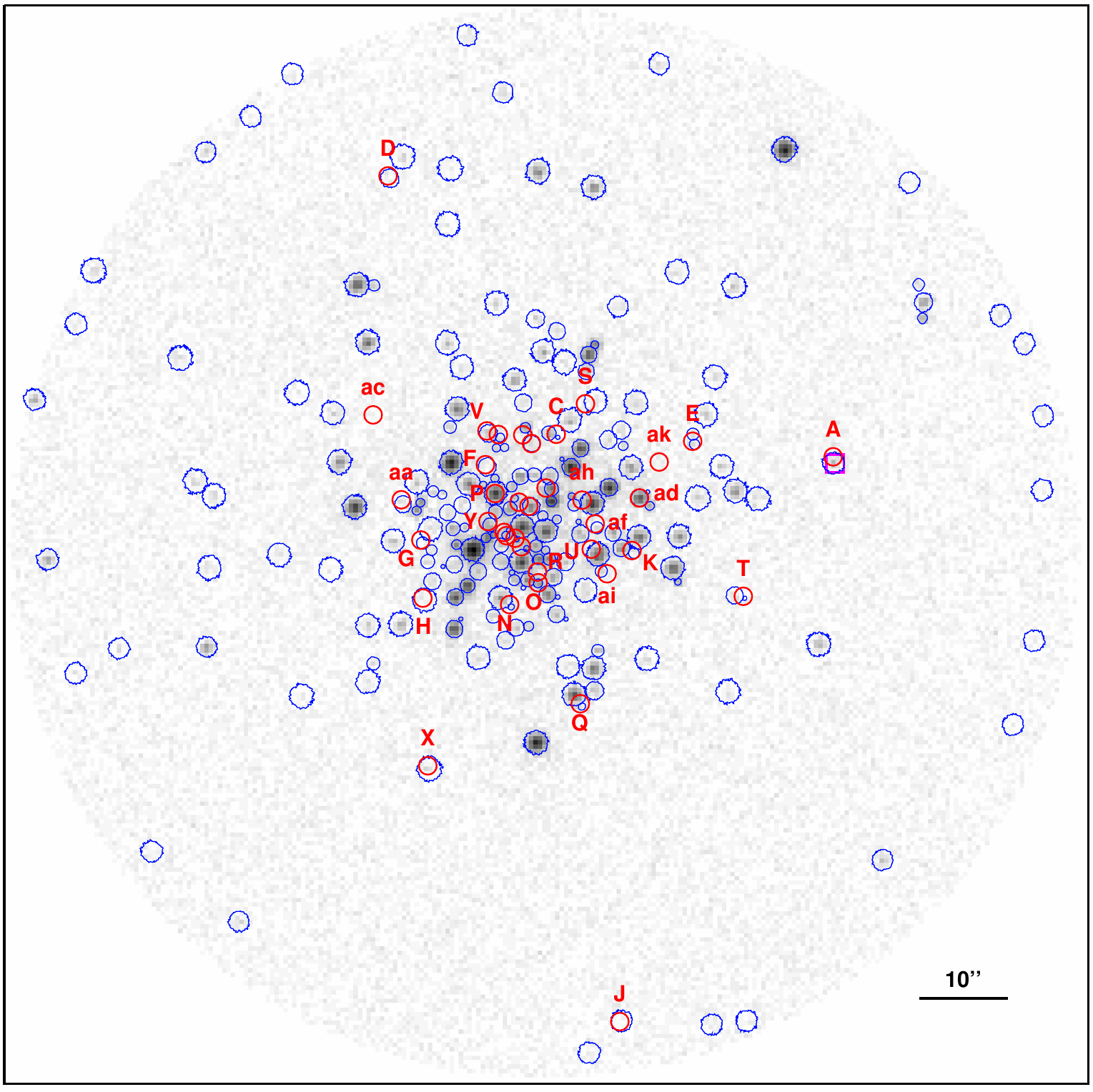}
\caption{Co-added \textit{Chandra} ACIS 0.5--10 keV image
    from 18 observations with a total effective exposure of 745.6\,ks of the core
    of Terzan 5. Exposures affected by luminous X-ray transient outbursts have not been  included.  The red circles of radius 1$''$ mark the radio positions of the 37 MSPs in the cluster with known positions. The X-ray sources detected within the cluster half-mass radius presented in \citet{Bahramian2020} are shown with the blue polygonal regions. For the MSP Ter5 A, the magenta square marks the Jansky VLA radio continuum position from \citet{Urquhart2020}. 
    The grey scale corresponds to number of counts per pixel increasing logarithmically from white to black.
  }
  \label{fig:image}
\end{figure*}

\begin{figure*}
\centering
\includegraphics[width=0.88\textwidth]{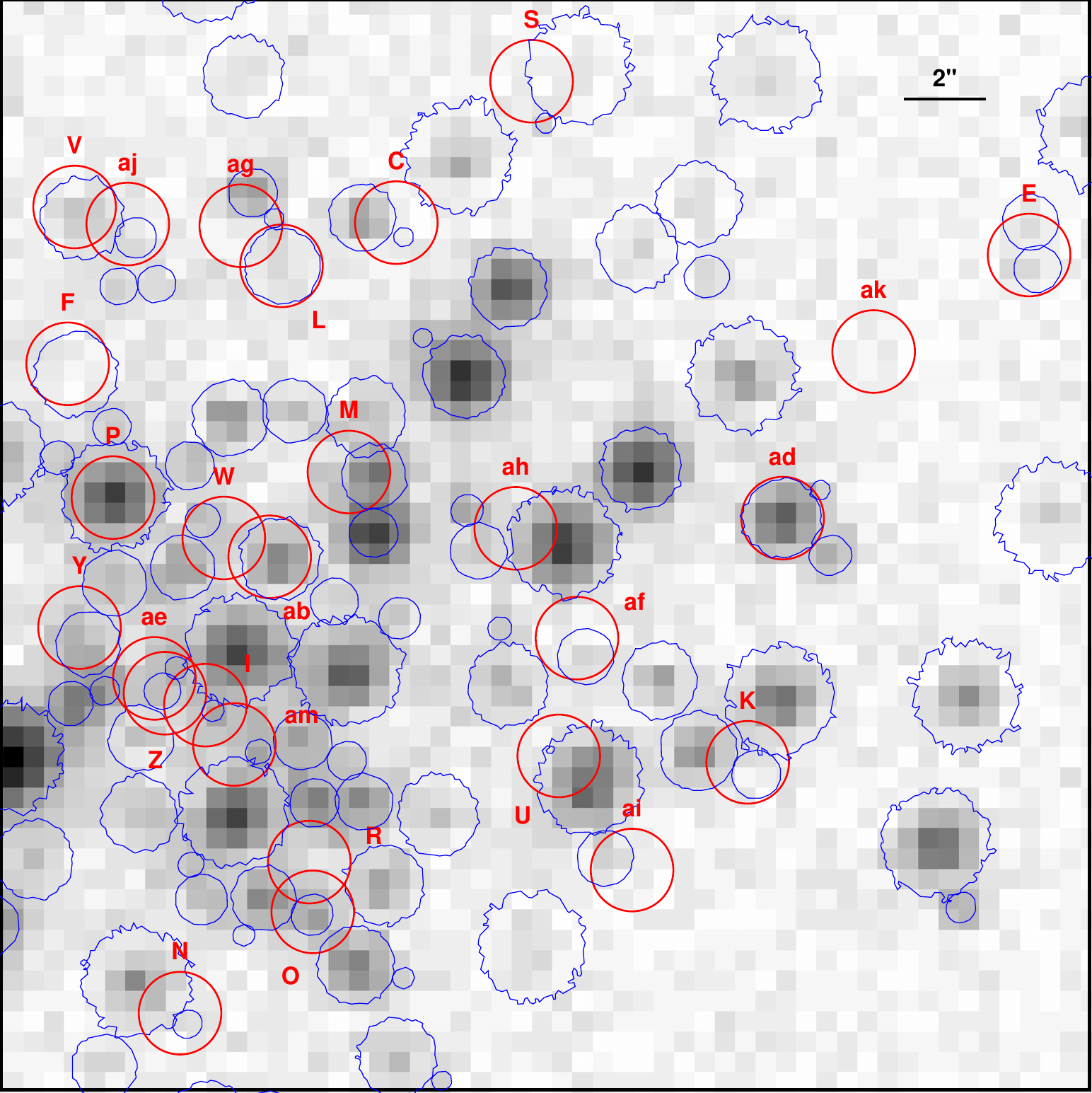}
\caption{Same as Figure~\ref{fig:image} but with a close-up of the crowded cluster core to show in more detail the MSP positions and X-ray sources.
  }
  \label{fig:images_zoom}
\end{figure*}

\section{Observations and Data Reduction}
\label{sec:obs}

Terzan 5 has been targeted with \textit{Chandra} on 23 separate occasions between 2000 and 2016. Table~\ref{tab:Obs} summarizes all available archival observations. Five of the observations (marked with *) are overwhelmed by emission from one of the cluster X-ray binaries in outburst so they are not used in this analysis. We reprocessed all \textit{Chandra} exposures of Terzan 5 with CIAO\footnote{Chandra Interactive Analysis of Observations \citep{2006SPIE.6270E..1VF}.} (4.10) and CALDB (4.8.0). First, we removed pixel randomization from the pipeline processing and applied the EDSER subpixel algorithm \citep{Li2004} to aid in disentangling the point sources in the crowded cluster core. To produce a stacked image, the relative astrometric offset of each observation was corrected by reprojecting the aspect solution using the coordinates of the brightest 40 sources scattered across the field of view.

The detailed source detection and parameter extraction procedures are described in \citet{Bahramian2020}. In brief, for X-ray source detection in the images including counterparts to the pulsars, we employed the CIAO tool \texttt{wavdetect}, which correlates the image with Mexican-hat wavelets over a range of scales to identify sources. We also applied the \texttt{PWDetect} script \citep{Damiani1997a,Damiani1997b}, which tends to be more effective at identifying faint sources near much brighter sources. Subsequently, we employed the IDL script \texttt{acis\_extract} \citep{Broos2010}, which provides a great deal of automation in the extraction process using the CIAO and FTOOLS software packages and is specifically designed for working with crowded fields and multiple observations. We used it to confirm the validity of the source detections reported by wavdetect and \texttt{PWDetect}, and refine the source positions. For MSPs where neither source detection algorithm identified a coincident X-ray source, we manually added a source at the position and fed it through the \texttt{acis\_extract} pipeline to extract the point source net count rates, spectra, light curves, and detection significance estimate of
each candidate source. 
The long frame time of the \textit{Chandra} ACIS exposures (3.2 seconds) does not allow a study of any pulsed X-ray emission from the MSPs in Terzan 5.

\begin{figure}
\centering
\includegraphics[width=0.46\textwidth]{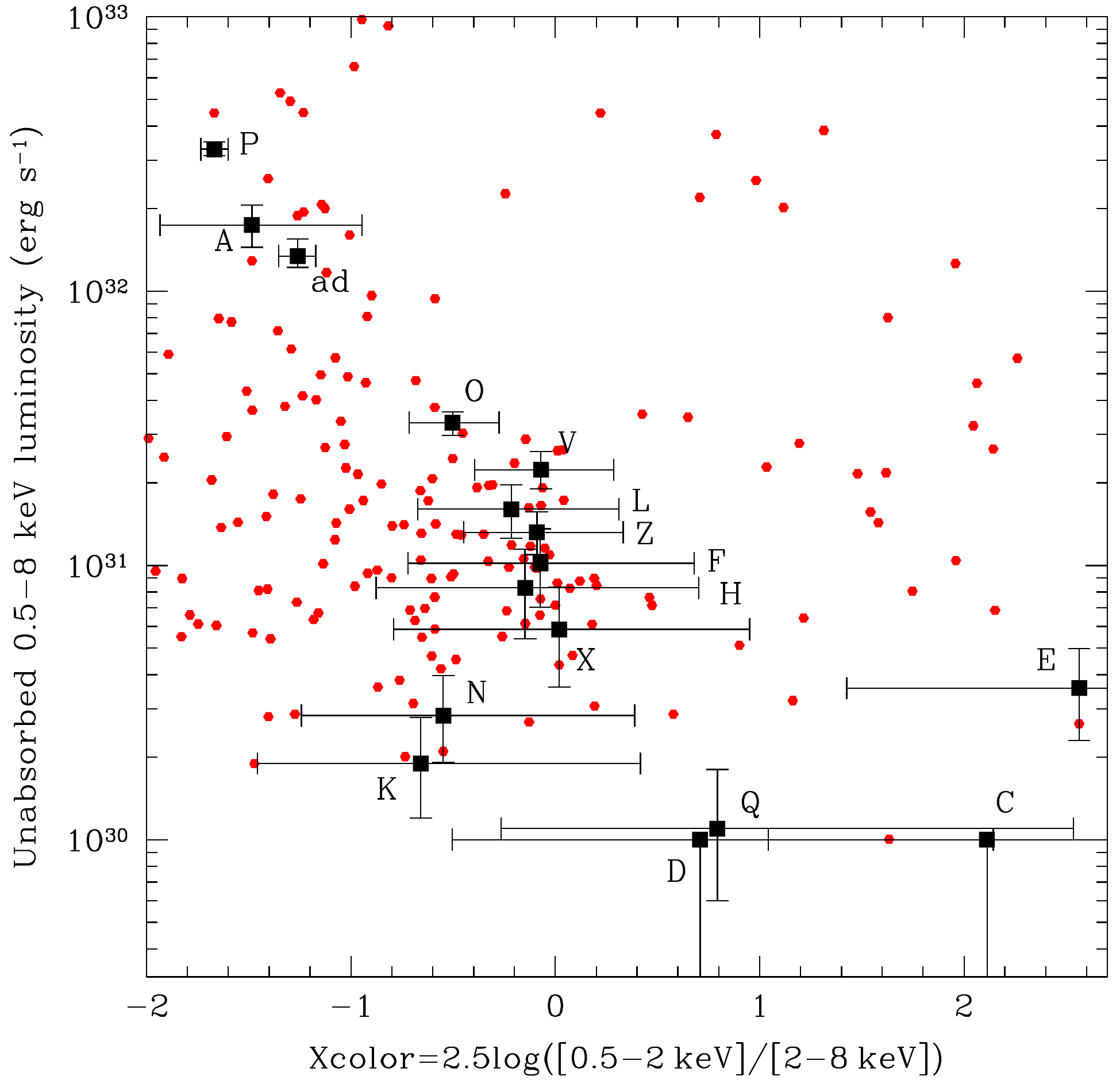}
\caption{X-ray color-luminosity diagram for Terzan 5. The red points mark all sources from the \citet{Bahramian2020} X-ray point source catalog, while the black squares correspond to the X-ray sources positionally coincident with the Terzan 5 radio MSPs. The labels in parentheses (for M, U, and ab) indicate that the association between the MSP and the coincident X-ray source is unlikely, while those marked with a question mark (for I and Y) indicate sources where the count and luminosity estimates may be unreliable due to source blending/crowding.
  }
  \label{fig:image_xcolor}
\end{figure}

\section{Imaging Analysis and X-ray Counterpart Matching}
\label{sec:imaging}

Figure~\ref{fig:image} show the \textit{Chandra} ACIS image of Terzan 5 produced by stacking all exposures not affected by bright transient outbursts, with an effective exposure of 745.6 ks. A zoom-in of the crowded cluster core is presented in Figure~\ref{fig:images_zoom} to reveal more clearly the overlapping source regions. Overlaid on the images are the radio timing positions of the 37 MSPs from \citet{Ransom05}, \citet{Andersen2018}, and \citet{Cadelano18}\footnote{The recently discovered Ter5 al has no reliable position determination so it is not included in this analysis.}, as well as the polygonal regions marking the X-ray sources from the catalog presented in \citet{Bahramian2020}.

To refine the astrometric alignment of the X-ray frame (the catalog of \citealt{Bahramian2020} uses 2MASS matches from \citealt{Heinke06}) with the MSP frame (based on the JPL planetary ephemeris), we first use the VLA interferometric position of the LMXB EXO 1745-248 \citep{Tetarenko16}, which is also present in the \citet{Bahramian2020} catalog. Adding offsets of $-0.18\arcsec$ in RA and $-0.31\arcsec$ in Dec to the X-ray positions matches these two positions of EXO 1745-248, each of which have statistical errors below $0.1\arcsec$. We apply this offset to all X-ray positions, and search for matches of radio timing MSP positions with the shifted X-ray positions, within the 95\% confidence error circles calculated using the empirical expression from \citet{Hong05}. 

Table~\ref{tab:srclist} summarizes the outcome of the imaging analysis and MSP X-ray source identification, showing the astrometric offset in right ascension ($\Delta\alpha$) and declination ($\Delta\delta$) between the radio MSP position and the nearest X-ray source, the  \citet{Hong05} 95\% confidence errors in the X-ray source position ($P_{\rm err}$) , and the ratio of the positional offset and the uncertainty in the X-ray position. We note that the radio MSP positions are typically determined to very high accuracy ($\ll 0.1\arcsec$) so the dominant uncertainty typically comes from the X-ray source positions (but see below for the case of Terzan 5 A).

We find 16 matches of MSPs with catalogued X-ray sources; and identify one match outside the 95\% confidence error circle (Ter5 A, discussed below). The catalogued X-ray sources come from different origins; some were detected blindly by detection algorithms, while others were inserted by hand at the locations of MSPs. These insertions were done in the cluster core due to severe crowding,\footnote{The problem is compounded by the high intervening column towards Terzan 5, which produces scattering halos that spread out the point source emission over a larger area on the detector.} which can make real sources hard for detection algorithms to identify (see Figure~\ref{fig:images_zoom}). For determining the number of real (versus spurious) matches, we treat these two groups separately, and also take into account the X-ray source quality classifications (poor detections have either false detection probability $>$1\% and/or $<$5 net counts) in \citet{Bahramian2020}. 

We find 8 MSPs that are high-confidence matches (MSPs identified with independently detected, good quality X-ray sources): P, ad, O, V, Z, L, E, and X. We identify the four MSPs K, N, F, and H as potential matches; the (good quality) X-ray sources here were added "by hand" to the source list as likely real sources at MSP positions. Although we are confident there are indeed X-ray sources at these positions, the identification of X-ray sources using MSP positions makes it difficult to assess the probability of a spurious match with these sources. Finally, we identify four MSPs that are possible matches to (poor quality) X-ray sources, C, D, J, and Q; these are also possibly real matches, but given their low detection significance, we are less confident in the reality of these X-ray sources.

As shown in Section~\ref{sec:orbital}, for P, O, and ad the associations can be considered as certain since the X-ray counterparts clearly show flux modulations at the binary orbital periods of the pulsars. These pulsars are also all excellent positional matches with the associated X-ray sources ($<0.1\arcsec$, or $<0.6\sigma$, offsets).
For the 8 independently detected X-ray sources matched with MSPs, we can assess the probability of a false match, by offsetting one source list and re-running the matching algorithm. This generally gives 2-3 spurious matches per run. Among the remaining sources, L, Z, and X match X-ray and radio positions within 1 sigma and $0.1\arcsec$, while V and E match at $0.3\arcsec$ and $1.6\sigma$ ($2\sigma$ is our cutoff for matches). Thus, we think L, Z, and X are most likely real matches, while V and E have a higher likelihood of being spurious matches. 

We address Terzan 5 A as a special case. The positional offset between the \citet{Prager2017} timing position and the X-ray position is $0.55\arcsec$, or $3\sigma$, so we would generally discount the match. However, Terzan 5 A is an eclipsing redback, of a class which often have timing solution difficulties due to the irregular eclipses throughout the orbit and changing dispersion measure \citep[see, e.g.,][]{Archibald13}. For instance, the derived timing solutions of the redbacks NGC 6397 A \citep{Bassa04} and M28 I \citep{Begin_2006,Bogdanov11,Papitto13} were in error by $0.811$ and $2\arcsec$, respectively. Terzan 5 A is also located well outside the core ($34\arcsec$, or 3.5 core radii, from the cluster center; see Fig.~\ref{fig:image}), where the probability of chance coincidence is greatly reduced, and redbacks are systematically X-ray brighter than other MSPs \citep{Roberts2014}, all of which make the association more plausible. 
We can independently test the position of Terzan 5 A using the deep Jansky VLA radio continuum observations of \citet{Urquhart2020}, where  Terzan 5 A is indeed detected. The VLA position of Terzan 5 A (17:48:02.249$\pm$0.05$\arcsec$, $-$24:46:37.65$\pm$0.10$\arcsec$) is indeed consistent with our (shifted) X-ray position (17:48:02.246$\pm$0.18$\arcsec$, $-$24:46:37.91$\pm$0.18$\arcsec$; errors quoted are 1$\sigma$), but neither is consistent with the radio timing position of \citet{Prager2017}, which does not state errors; (17:48:02.251, $-$24:46:37.37). The VLA position of Terzan 5 A is $0.28\arcsec$ from its radio timing position, while all other VLA positions of detected Terzan 5 pulsars match the timing positions exquisitely ($<0.05\arcsec$). As the flux density of the VLA source matches Terzan 5 A's pulsed flux density, and the radio spectral index ($\approx -3$) matches expectations of a radio pulsar, we conclude that the VLA position for Terzan 5 A is more accurate, and that we have probably detected the true X-ray counterpart of Terzan 5 A.

For the possible source matches, we can further test  their plausibility by examining the source properties and comparing them against expectations of the X-ray properties of MSPs. For this purpose, we construct a color-magnitude diagram following \citet{Heinke06}, with the color expressed as $2.5\log{[0.5-2\,{\rm keV}]/[2-8\,{\rm keV}]}$ and the estimated unabsorbed luminosity in the 0.5--8 keV range from Table 4. The latter was estimated assuming a power-law spectrum with photon index $\Gamma=2$, which is a reasonable approximation for a variety of X-ray spectral types including non-thermal emission, thermal plasmas, and the higher energy portion ($\gtrsim0.5$\,keV) of the soft spectra of MSPs. Figure~\ref{fig:image_xcolor} shows the X-ray color-magnitude diagram of Terzan 5 and includes all sources detected within the half-mass radius of Terzan 5 (red circles), with the X-ray source matches to the MSP highlighted (black squares). 
The X-ray counterparts of the three redback MSP systems, Ter5 A, P, and ad clearly stand out from the sample with luminosities $\gtrsim 10^{32}$\,erg\,s$^{-1}$ and hard spectral colors, consistent with the observed properties of the current sample of such systems in other clusters and the field of the Galaxy (see \ref{sec:conclusions} for a discussion). The counterpart of the black widow Ter5 O shows a moderately high luminosity and hardness, comparable to those of the canonical member of this binary MSP class, PSR B1957+20 \citep[see][]{Huang2012}. Ter5 ae is also a black widow MSP \citep{Prager2017}, but it partially overlaps with Ter5 Z (see Figure~\ref{fig:images_zoom}) and is situated in the wings of a much brighter source (CX9 from \citealt{Heinke06}). Thus, it is difficult to definitively attribute the observed X-ray emission to this pulsar, especially since the councident X-ray emission shows no statistically significant evidence of orbital variability.  The majority of X-ray counterparts tend to be softer and relatively faint, with $L_X \lesssim 10^{31}$\,erg\,s$^{-1}$, with relatively soft spectra. Although these properties are generally consistent with the X-ray characteristics of most MSPs, it is not possible to conclusively establish an association between the radio MSP and the superposed X-ray source due to a high likelihood of chance coincidences and the absence of other indicators of a connection.  For MSPs Ter5 J, ac, and ak, which are situated on the periphery of the crowded cluster core, there is no statistically significant excess of counts above the background and can thus be considered as conclusive non-detections.

\begin{deluxetable*}{lclcrrrrrrRRRR}
\tablecolumns{14} 
\tablewidth{0pt}  
\tabletypesize{\scriptsize}
\tablecaption{Summary of X-ray Properties of the Terzan 5 MSPs \label{tab:srclist}}  
\tablehead{\colhead{PSR} &  \colhead{Source} &  \colhead{CXOU J} & \colhead{Right ascension} & \colhead{Declination} &  \colhead{$P_{\rm err}$} & \colhead{$\Delta\alpha$} & \colhead{$\Delta\delta$}  & $\Delta/P_{\rm err}$ & \colhead{Exposure} & \colhead{Net Counts} & \colhead{Net Counts} &  \colhead{$\sigma_X$} & \colhead{$L_X$\tablenotemark{b}}  \\ 
            \colhead{}      & \colhead{label}      & \colhead{name\tablenotemark{a}}  &  \colhead{(hh:mm:ss.ss)} & \colhead{(dd:mm:ss.ss)} & \colhead{($\arcsec$)} & \colhead{($\arcsec$)} &     \colhead{($\arcsec$)}     & \colhead{ } &  \colhead{(ks)} &  \colhead{(0.5--2 keV)} & \colhead{(2--8 keV)} & \colhead{ } &  \colhead{($10^{31}$\,erg\,s$^{-1}$)}  }
\startdata
\textbf{A}	&	67	&	\textbf{174802.26--244637.5}	&	  17:48:02.224  &   -24:46:37.91   & 0.36	&	0.11	&	0.54	&	1.53	&	42.5	&	6.6_{-2.4}^{+3.0}	&	25.9_{-4.9}^{+5.4}	&	4.8	&	8.6\pm2.6	\\
C	&	172	&	174804.53--244634.7\tablenotemark{c}
&	17:48:04.517	& -24:46:35.17	&	0.48	& 0.26	&	0.45	&	1.08	&	738.2	&	$<$2.8		&	$<$0.8		&	0.1	&	<0.08	\\
D	&	173	&	174805.92--244605.6\tablenotemark{c}  & 17:48:05.926 & -24:46:06.05
&	0.48	&	-0.05	&	0.38	&	0.80	&	738.2	&	4.8	_{-2.8}^{+3.3}	&	2.5_{-2.4}^{+2.8}	&	1.3	&	<0.1	\\
E	&	174	&	174803.40--244635.4  &  17:48:03.413    &   -24:46:35.78
&	0.42	&	-0.06	&	0.30	&	0.72	&	738.2	&	10.6_{-3.3}^{+3.8}	&	1_{-1.0}^{+2.7}	&	2.1	&	0.26_{-0.09}^{+0.11}	\\
F	&	175	&	174805.11--244638.0\tablenotemark{c}
 &  17:48:05.110 &  -24:46:38.16 
&	0.40	&	0.13	&	0.11	&	0.43	&	332.5	&	7.2_{-3.0}^{+3.5}	&	7.7_{-3.3}^{+3.8}	&	2.5	&	0.76_{-0.23}^{+0.25}	\\
G	&	\nodata	&	\nodata	&	\nodata	&	\nodata	&	\nodata	& \nodata		&	\nodata	&	\nodata	&	\nodata	&	\nodata		&	\nodata		&	\nodata	 &	\nodata	\\
H	&	177	&	174805.63--244653.0\tablenotemark{c} &  17:48:05.621    &   -24:46:53.24
&	0.41	&	0.20	&	0.18	&	0.66	&	364.7	&	6.2_{-3.0}^{+3.5}	&	7.1	_{-3.3}^{+3.8}	&	2.3	&	0.61_{-0.21}^{+0.24}	\\
I &	\nodata	&	\nodata	&	\nodata	&	\nodata	&	\nodata	& \nodata		&	\nodata	&	\nodata	&	\nodata	&	\nodata		&	\nodata		&	\nodata	 &	\nodata	\\	
J &	\nodata	&	\nodata	&	\nodata	&	\nodata	&	\nodata	& \nodata		&	\nodata	&	\nodata	&	\nodata	&	\nodata		&	\nodata		&	\nodata	 &	\nodata	\\
K	&	179	&	174803.90--244647.7\tablenotemark{c}
 &  17:48:03.900    &   -24:46:47.84
&	0.45	&	0.14	&	0.11	&	0.40	&	745.6	&	3_{-1.7}^{+2.4}	&	5.5_{-2.6}^{+3.2}	&	1.9	&	0.19_{-0.07}^{+0.09}	\\
\textbf{L}	&	136	&	\textbf{174804.75--244635.4}	&
 17:48:04.735   &   -24:46:35.75  &
0.37	&	0.04	&	-0.06	&	0.19	&	347.6	&	11_{-3.4}^{+4.0}	&	13.4_{-3.8}^{+4.3}	&	3.8	&	1.19_{-0.26}^{+0.27}	\\
M	&	\nodata	&	\nodata	&	\nodata	&	\nodata	&	\nodata	& \nodata		&	\nodata	&	\nodata	&	\nodata	&	\nodata		&	\nodata		&	\nodata	 &	\nodata	\\
N	&	180	&	174804.91--244653.7\tablenotemark{c} &  17:48:04.906    &   -24:46:54.00
&	0.45	&	0.20	&	0.18	&	0.60	&	745.6	&	3.5_{-1.7}^{+2.4}	&	5.8_{-2.4}^{+3.0}	&	2.1	&	0.21_{-0.07}^{+0.08}	\\
\textbf{O}	&	18	&	\textbf{174804.69--244651.1}	&
 17:48:04.687   &   -24:46:51.41    &
0.32	&	-0.08	&	0.01	&	0.25	&	738.2	&	41.5_{-6.3}^{+6.9}	&	65.9_{-8.2}^{+8.4}	&	9.3	&	2.45_{-0.24}^{+0.24}	\\
\textbf{P}	&	54	&	\textbf{174805.05--244641.0}	&
 17:48:05.038   &   -24:46:41.40    & 
0.29	&	0.01	&	0.04	&	0.14	&	745.6	&	349.6_{-19.6}^{+19.4}	&	162_{-36.0}^{+44.0}	&	43.0	&	32.9^{+2.1}_{-1.8}	\\
Q	&	181	&	174804.33--244704.9\tablenotemark{c}	
 &  17:48:04.332    &   -24:47:05.12 
&	0.45	&	0.06	&	0.09	&	0.24	&	738.2	&	2.7_{-1.5}^{+2.2}	&	1.3_{-1.1}^{+1.8}	&	1.2	&	0.11_{-0.05}^{+0.07}	\\
R	&	\nodata	&	\nodata	&	\nodata	&	\nodata	&	\nodata	& \nodata		&	\nodata	&	\nodata	&	\nodata	&	\nodata		&	\nodata		&	\nodata	 &	\nodata	\\
S	&	\nodata	&	\nodata	&	\nodata	&	\nodata	&	\nodata	& \nodata		&	\nodata	&	\nodata	&	\nodata	&	\nodata		&	\nodata		&	\nodata	 &	\nodata	\\
T	&	\nodata	&	\nodata	&	\nodata	&	\nodata	&	\nodata	& \nodata		&	\nodata	&	\nodata	&	\nodata	&	\nodata		&	\nodata		&	\nodata	 &	\nodata	\\
U	&	\nodata	&	\nodata	&	\nodata	&	\nodata	&	\nodata	& \nodata		&	\nodata	&	\nodata	&	\nodata	&	\nodata		&	\nodata		&	\nodata	 &	\nodata	\\
V & 134 & 174805.10--244634.2 &  17:48:05.090    &   -24:46:34.63
&	0.35	&	0.23	&	0.17	&	0.82	&	475.5	&	22.6_{-4.8}^{+5.3}	&	24.1_{-5.1}^{+5.5}	&	5.6	&	1.65_{-0.25}^{+0.27}	\\
W	&	\nodata	&	\nodata	&	\nodata	&	\nodata	&	\nodata	& \nodata		&	\nodata	&	\nodata	&	\nodata	&	\nodata		&	\nodata		&	\nodata	 &	\nodata	\\
\textbf{X}	&	183	&	\textbf{174805.59--244712.0}	&	
 17:48:05.587   &   -24:47:12.14    & 
0.42	&	0.10	&	0.08	&	0.30	&	446.5	&	5.8	_{-2.9}^{+3.3}	&	5.7	_{-3.2}^{+3.7}	&	2.0	&	0.43_{-0.17}^{+0.18}	\\
Y	&	\nodata	&	\nodata	&	\nodata	&	\nodata	&	\nodata	& \nodata		&	\nodata	&	\nodata	&	\nodata	&	\nodata		&	\nodata		&	\nodata	 &	\nodata	\\
\textbf{Z}	&	186	&	\textbf{174804.96--244645.7}
 &  17:48:04.951    &   -24:46:46.04 
&	0.36	&	-0.08	&	-0.07	&	0.30	&	584.0	&	16.2_{-3.8}^{+4.5}	&	17.6_{-4.1}^{+4.8}	&	4.8	&	0.98_{-0.16}^{+0.18}	\\
aa	&	\nodata	&	\nodata	&	\nodata	&	\nodata	&	\nodata	& \nodata		&	\nodata	&	\nodata	&	\nodata	&	\nodata		&	\nodata		&	\nodata	 &	\nodata	\\
ab	&	\nodata	&	\nodata	&	\nodata	&	\nodata	&	\nodata	& \nodata		&	\nodata	&	\nodata	&	\nodata	&	\nodata		&	\nodata		&	\nodata	 &	\nodata	\\
ac	&	\nodata	&	\nodata	&	\nodata	&	\nodata	&	\nodata	& \nodata		&	\nodata	&	\nodata	&	\nodata	&	\nodata		&	\nodata		&	\nodata	 &	\nodata	\\
\textbf{ad}	&	52	&	\textbf{174803.86--244641.5}	&
 17:48:03.850   &   -24:46:41.94  & 
0.30	&	-0.04	&	0.08	&	0.30	&	745.6	&	198.6_{-14.6}^{+14.4}	&	634.5_{-25.5}^{+25.5}	&	27.6	&	13.4^{+1.6}_{-1.2}	\\
ae	&	\nodata	&	\nodata	&	\nodata	&	\nodata	&	\nodata	& \nodata		&	\nodata	&	\nodata	&	\nodata	&	\nodata		&	\nodata		&	\nodata	 &	\nodata	\\
af	&	\nodata	&	\nodata	&	\nodata	&	\nodata	&	\nodata	& \nodata		&	\nodata	&	\nodata	&	\nodata	&	\nodata		&	\nodata		&	\nodata	 &	\nodata	\\
ag	&	\nodata	&	\nodata	&	\nodata	&	\nodata	&	\nodata	& \nodata		&	\nodata	&	\nodata	&	\nodata	&	\nodata		&	\nodata		&	\nodata	 &	\nodata	\\
ah	&	\nodata	&	\nodata	&	\nodata	&	\nodata	&	\nodata	& \nodata		&	\nodata	&	\nodata	&	\nodata	&	\nodata		&	\nodata		&	\nodata	 &	\nodata	\\
ai	&	\nodata	&	\nodata	&	\nodata	&	\nodata	&	\nodata	& \nodata		&	\nodata	&	\nodata	&	\nodata	&	\nodata		&	\nodata		&	\nodata	 &	\nodata	\\
aj	&	\nodata	&	\nodata	&	\nodata	&	\nodata	&	\nodata	& \nodata		&	\nodata	&	\nodata	&	\nodata	&	\nodata		&	\nodata		&	\nodata	 &	\nodata	\\
ak	&	\nodata	&	\nodata	&	\nodata	&	\nodata	&	\nodata	& \nodata		&	\nodata	&	\nodata	&	\nodata	&	\nodata		&	\nodata		&	\nodata	 &	\nodata	\\
am	&	\nodata	&	\nodata	&	\nodata	&	\nodata	&	\nodata	& \nodata		&	\nodata	&	\nodata	&	\nodata	&	\nodata		&	\nodata		&	\nodata	 &	\nodata	\\
\enddata
\tablenotetext{a}{Astrometric matches of Terzan 5 {\it Chandra} point source catalog \citep{Bahramian2020} to radio MSP position. Entries in bold face are X-ray source matches we deem as secure. The MSPs with no information listed have no statistically significant X-ray source situated within a 95\% confidence matching radius. RA and Dec are the X-ray positions, shifted to the VLA frame (a shift of -0.18" in RA, -0.31" in Dec, compared to \citealt{Bahramian2020}).}
\tablenotetext{b}{Unabsorbed X-ray luminosity in the 0.5--8 keV range in units of $10^{31}$ erg s$^{-1}$ assuming a distance to Terzan 5 of 5.9\,kpc and $N_{\rm}=2.1\times10^{22}$\,cm$^{-2}$. For A, P, and ad, $L_X$ was derived from a spectroscopic analysis. For the rest, $L_X$ was estimated assuming a power-law spectrum with photon index $\Gamma=2$.}
\tablenotetext{c}{X-ray source added manually at pulsar position.}
\end{deluxetable*}

\begin{figure}
\includegraphics[clip,trim = 0.7cm 10.0cm 1.2cm 2cm,width=0.45\textwidth]{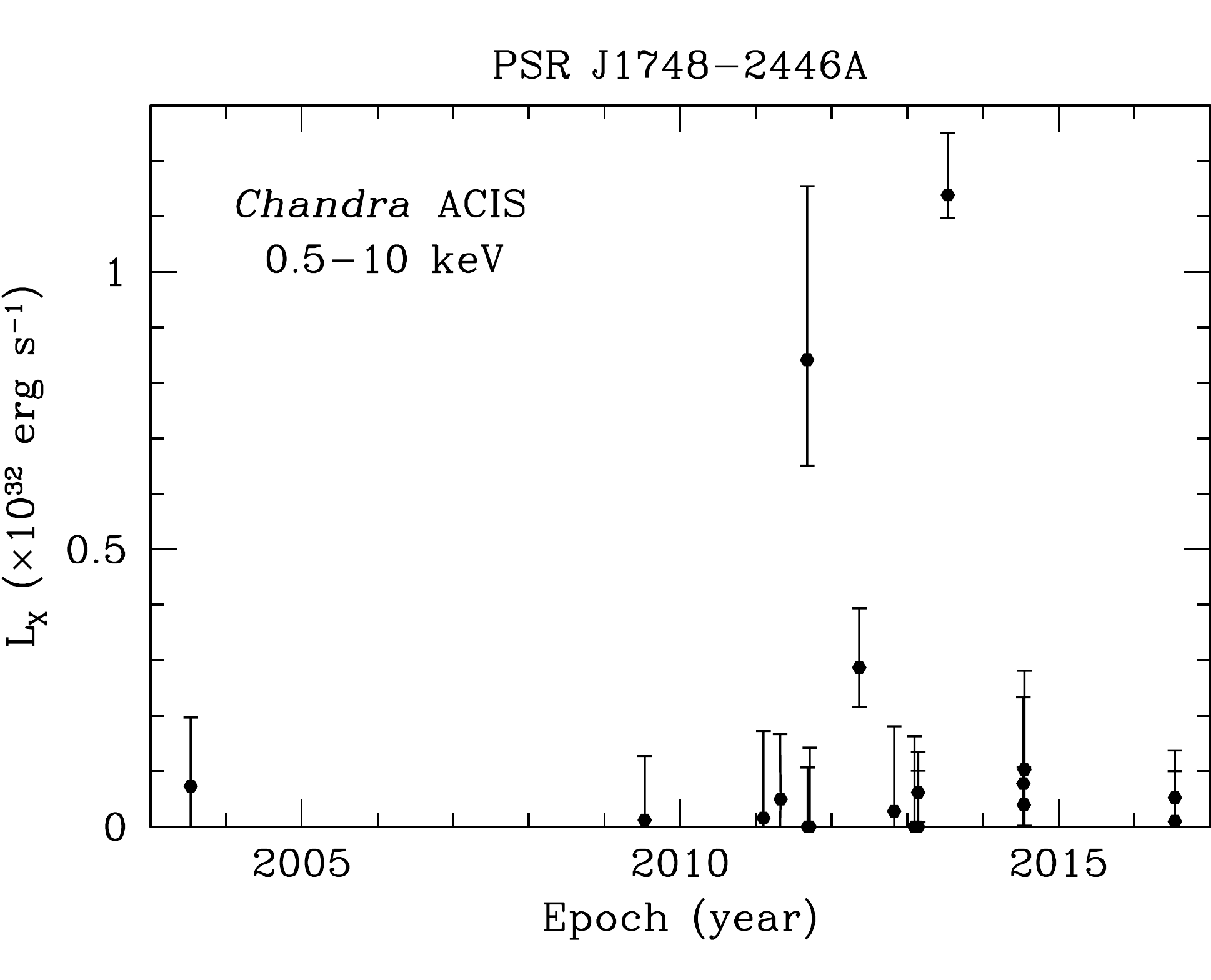}
\includegraphics[width=0.46\textwidth]{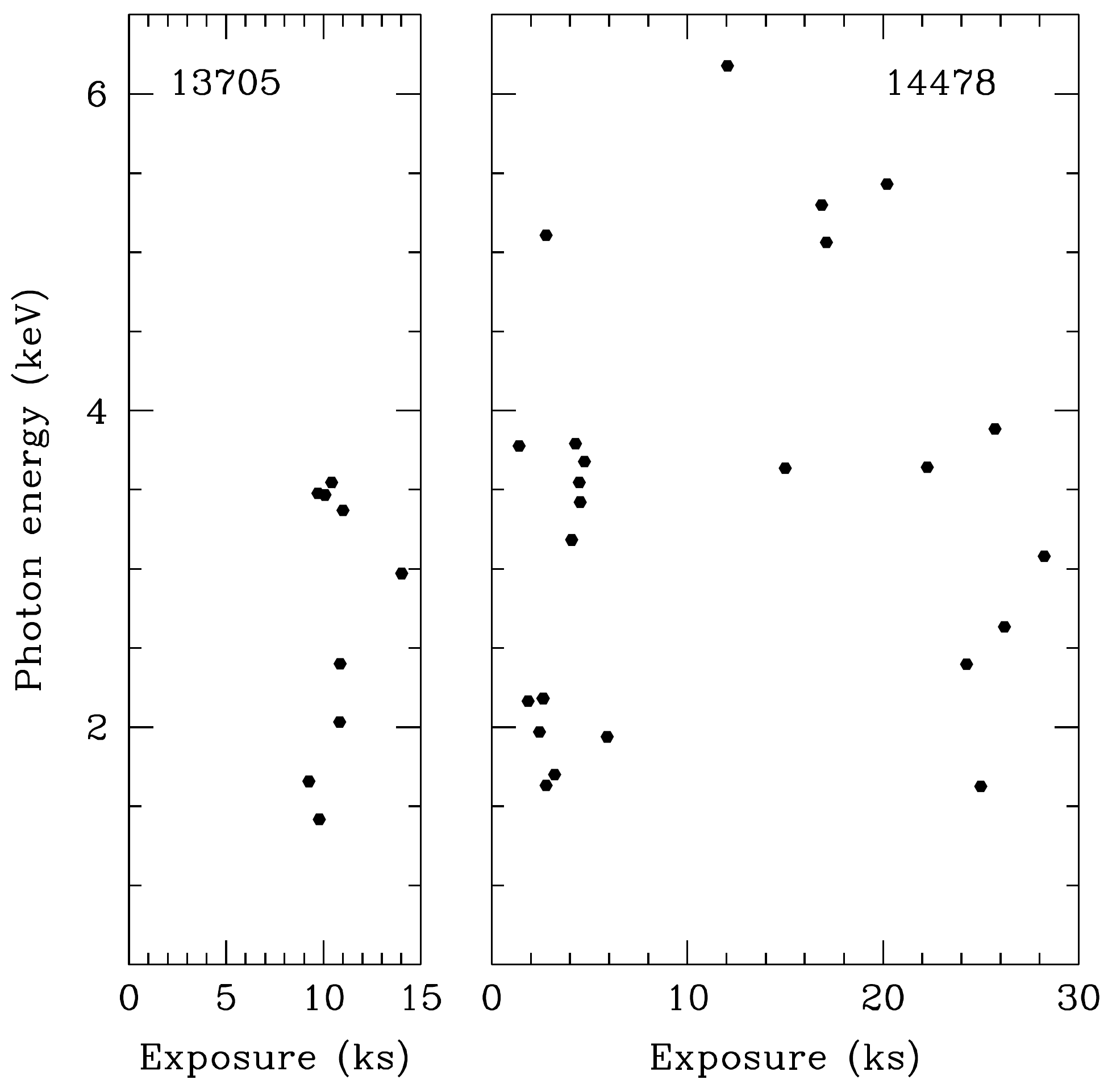}
\caption{Top: \textit{Chandra} ACIS 0.5--10\,keV long term background-subtracted light curve (unabsorbed X-ray luminosity versus time) of the X-ray counterpart of Ter5 A. In most exposures, the source is at the limit of detection, except on  2011 September 5 and  2013 July 16, when the X-ray luminosity is  $\approx 1\times 10^{32}$ erg s$^{-1}$. Bottom: Photon energy versus exposure time for the events detected in the Ter5 A X-ray counterpart source aperture during ObsIDs 13705 and 14478. The clustering in time of events is suggestive of flaring/bursting activity.}
  \label{fig:ter5a_lcurve}
\end{figure}

\section{Long-term Flux Variability}
\label{sec:variability}

Given the recent discovery of transitional MSPs (tMSPs), which appear to be drawn from a parent population of redback binaries, it is of interest to examine the variability of the three Terzan 5 redback MSPs systems on time scales of years. Based on the cases identified so far, a state transition from a radio pulsar state to an accretion-disk dominated state is accompanied by an increase in X-ray luminosity by roughly an order of magnitude to a level of a few $\times10^{33}$\,erg\,s$^{-1}$ \citep[see, e.g.,][]{Stappers2014,Bassa14}. In the accretion-disk dominated state, the X-ray emission exhibits a characteristic variability pattern with sudden switches between two well defined flux levels \citep{Linares14,deMartino2013,Bogdanov2015}, interspersed with flares reaching $L_X\times10^{34}$\,erg\,s$^{-1}$ . \citet{Bahramian2018} found that the Terzan 5 X-ray source CX1 exhibits X-ray and radio behavior on short and long timescales consistent with those of confirmed tMSPs. With the 16 years of coverage with \textit{Chandra}, we can look for evidence of state transitions in the known binary MSPs in the cluster.

The redback binaries Ter5 P and ad, and the black widow Ter5 O show marginal evidence for flux variations in their full time series, which can be attributed to orbital variability, as shown in the following section. The redback Ter5 A has an average count rate that is over an order of magnitude lower compared to the other two redbacks. Moreover, 32 of the 79 total counts extracted from the source aperture occur in two fairly short exposures on 2011 September 5 (ObsID 13705, 13.9 ks) and 2013 Jul 16 (ObsID 14478, 28.6 ks). The average luminosity of Ter5 A for these observations is $\approx 1\times 10^{32}$\,erg\,s$^{-1}$ (0.5--8 kev), whereas the time-averaged flux over the remaining  exposure time is below $\sim 1\times10^{31}$\,erg\,s$^{-1}$ in the same energy range (see Figure~\ref{fig:ter5a_lcurve}. A close inspection of the distribution in time of the events coincident with Ter5 A from ObsIDs 13705 and 14478, reveals that most of them arrive during narrow windows of $\sim$1 and $\sim$5\,ks (see bottom panels of Figure~\ref{fig:ter5a_lcurve}), indicative of short-lived bursting or flaring activity reaching $L_X\approx 1\times 10^{33}$\,erg\,s$^{-1}$. 

This large-amplitude change in the source luminosity suggests that Ter5 A experienced a state transition to an accretion disk dominated state twice. The peak luminosity in the short flares is comparable to the level seen from other transitional systems, including CX1.  An alternative interpretation for the variability is enhanced X-ray luminosity produced by intense stellar flaring from the secondary star as is seen from redbacks in the field of the Galaxy \citep{Cho18,Halpern17}. In the absence of contemporaneous optical and radio data it is difficult to establish the correct scenario, although the latter appears more plausible.

\section{X-ray Binary Orbital Variability}
\label{sec:orbital}

The MSPs known to exhibit radio eclipses are of additional interest as the X-ray emission of such systems is often  dominated by synchrotron radiation from the shock formed by the interaction of the pulsar wind with matter from the companion. This emission shows orbital-phase-dependent variability, first identified in the globular cluster redback systems 47 Tuc W \citep{Bogdanov05}, NGC 6397A \citep{Bogdanov10}, and M28H \citep{Bogdanov11}, as well as a growing number of redbacks in the field of the Galaxy \citep[see][and references therein]{Bogdanov2017}. 

For the known redback MSPs Ter5 P and ad, and the two black widow systems Ter5 O and ae, we can look for any X-ray modulation by folding the X-ray photons at the binary periods of these pulsars determined from radio timing ($P_b=8.7$, $26.3$, $6.2$, and $4.1$\,h,  respectively)\footnote{Redback binaries are known to exhibit non-deterministic variations in the orbital period due to tidal interactions. However, these are typically at the level of seconds to minutes and thus do not significantly affect the outcome of our analysis.}. The bins of the resulting count rate X-ray light curves were weighted to correct for the non-uniform orbital phase coverage of the observations. To determine the statistical significance of the orbital modulation, we employ the Kuiper test \citep{Paltani2004}, which is suitable for discontinuous time series and makes use of the unbinned event list. 

\begin{figure}
\centering
\includegraphics[clip,trim = 0.8cm 2.3cm 0cm 0cm,width=0.44\textwidth]{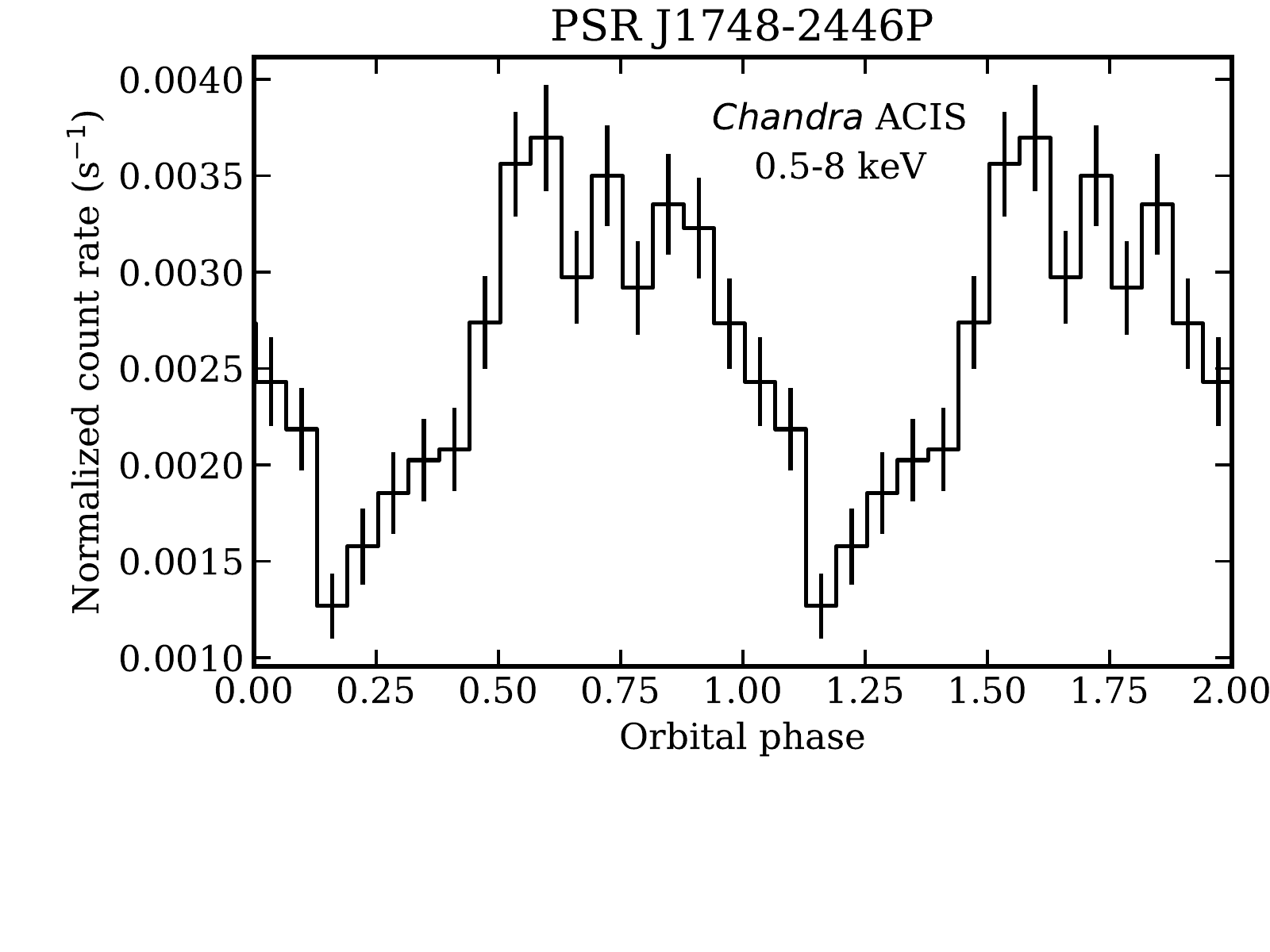}
\includegraphics[clip,trim = 0.8cm 2.3cm 0cm 0cm,width=0.44\textwidth]{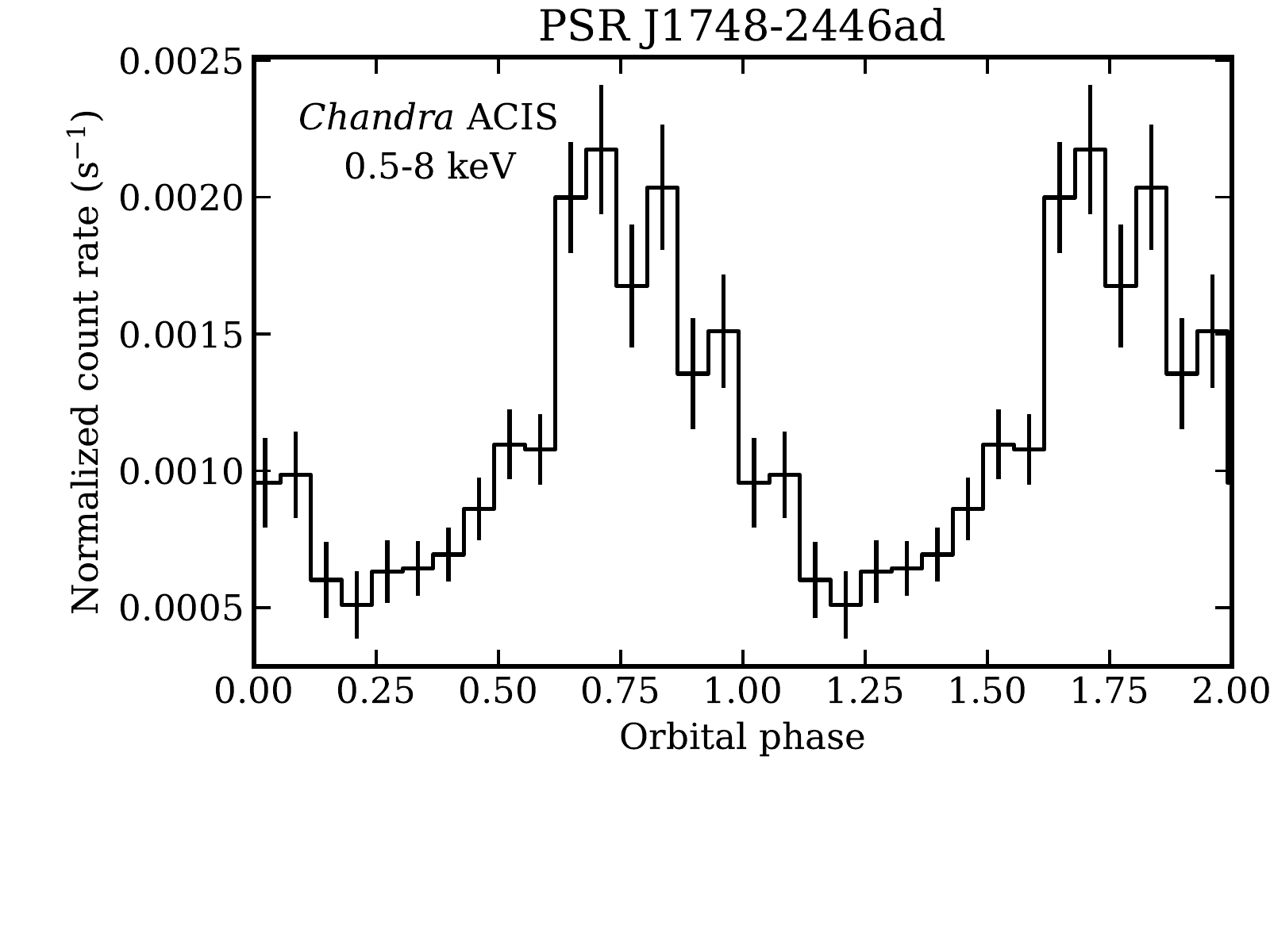}
\includegraphics[clip,trim = 0.8cm 2.3cm 0cm 0cm,width=0.44\textwidth]{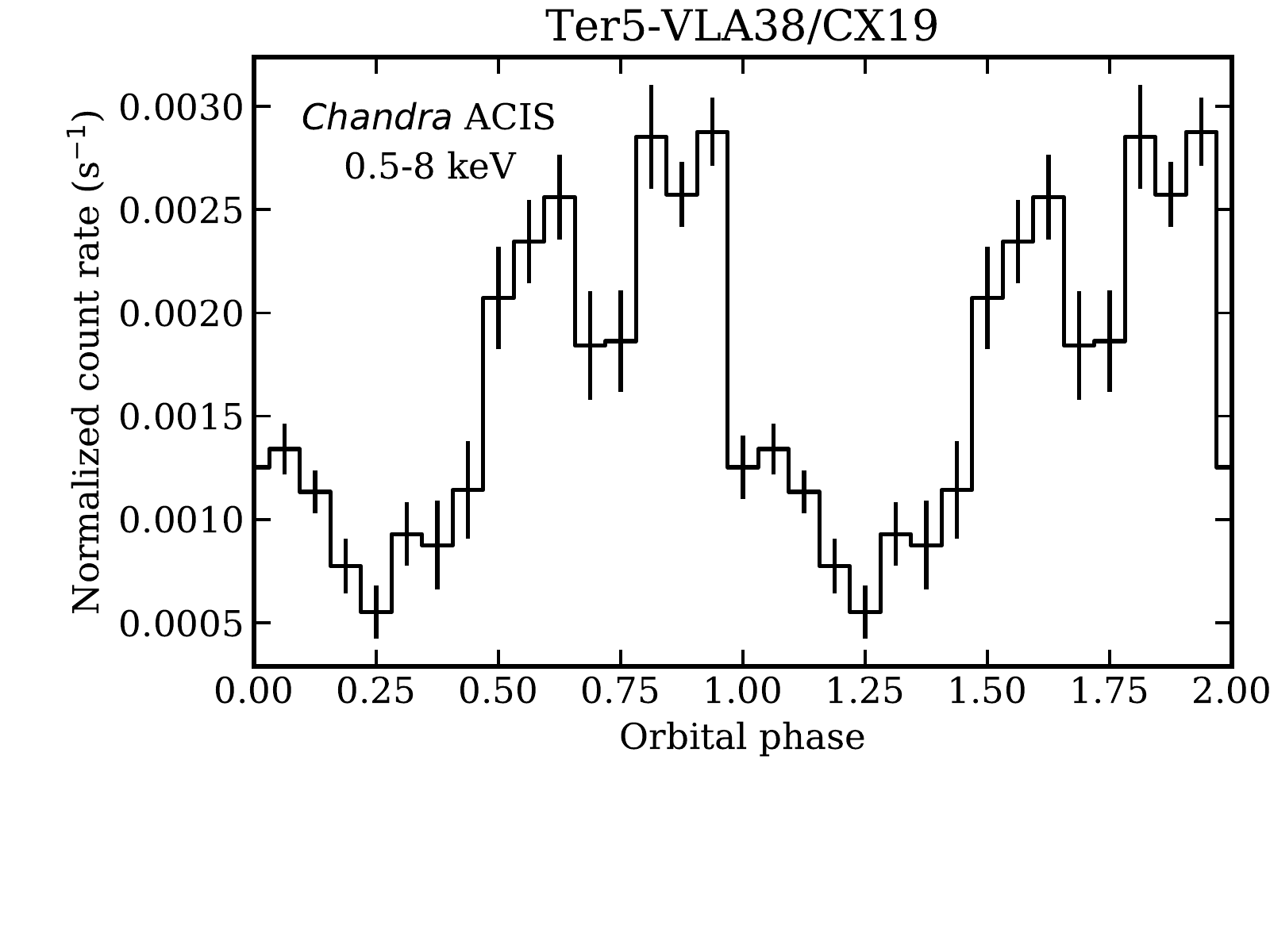}
\caption{\textit{Chandra} ACIS 0.5--8\,keV exposure-corrected, binned light curves of the redback MSPs, Ter5 P (PSR J1748--2446P) and Ter5 ad (PSR J1748--2446ad), folded at their respective binary periods. For an illustrative comparison, we also show the folded light curve of the X-ray counterpart of the redback candidate Ter5-VLA38 presented in \citet{Urquhart2020}, with arbitrary orbital phase. Two orbital cycles are shown for clarity.
  }
  \label{fig:lcurves}
\end{figure}

The exposure-corrected light curves for Ter5 P and Ter5 ad are shown in Figure~\ref{fig:lcurves}, for which we find probabilities of $2.7\times10^{-27}$ ($10.8\sigma$) and $2.4\times10^{-23}$ ($9.9\sigma$) that photons drawn from a uniform distribution in orbital phase would exhibit this level of non-uniformity. For reference, we also show the folded light curve of the redback candidate Ter5-VLA38 identified by \citet{Urquhart2020} to be coincident with X-ray source CX19 from \citet{Heinke06}, to highlight the close similarities in the variability pattern seen in these systems. The X-ray minima for both systems occur at superior conjunction ($\phi\approx 0.25$), when the secondary star is situated between the pulsar and observer.

\begin{figure}
\centering

\includegraphics[clip,trim = 0.8cm 2.4cm 0cm 0cm,width=0.44\textwidth]{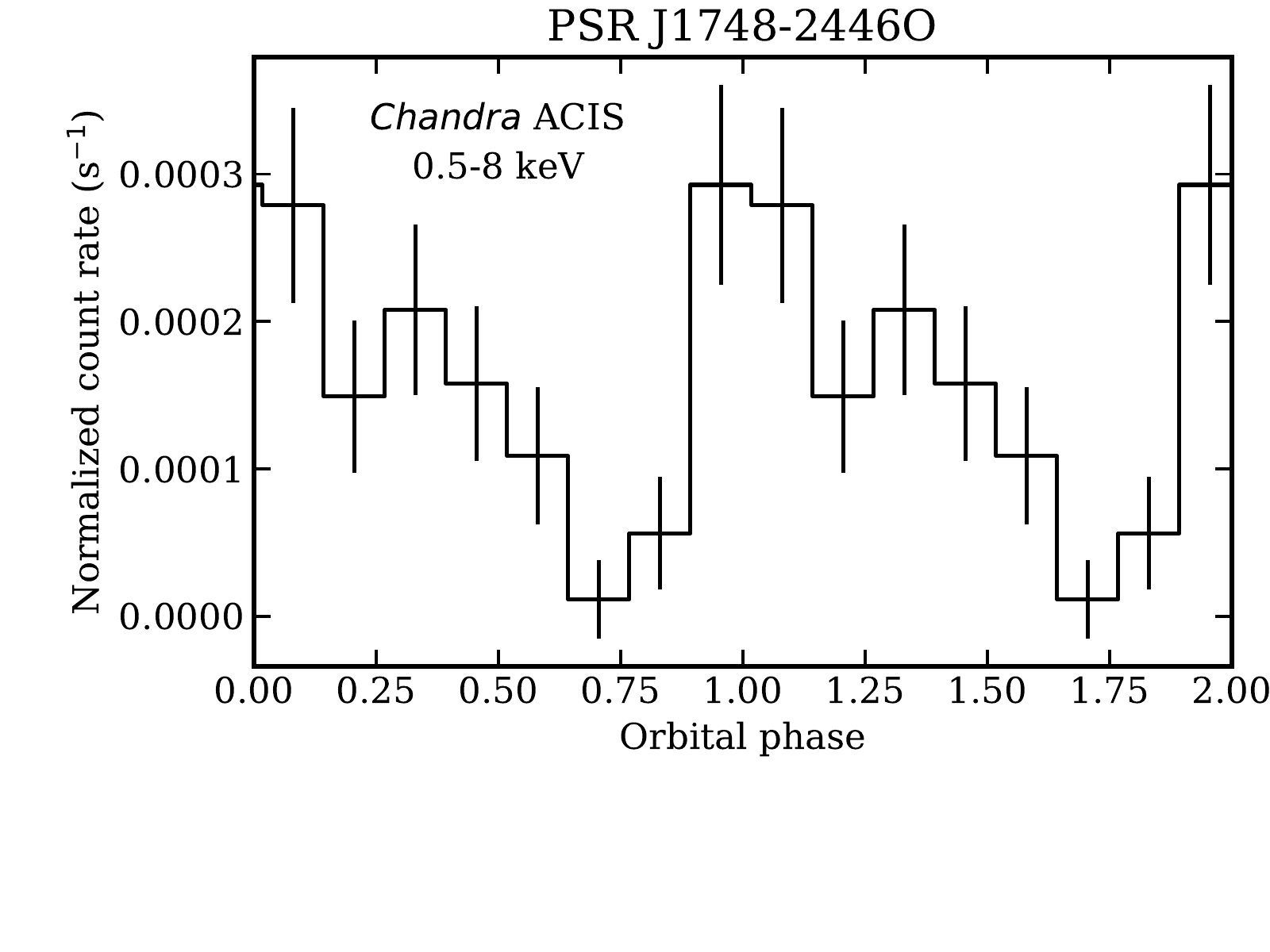}

\caption{Same as Figure~\ref{fig:lcurves} but for the black widow binary MSP Ter5 O (PSR J1748--2446O). The count rate in the lowest bin at $\phi_b \approx0.75$ is consistent with zero.
  }
  \label{fig:lcurvebw}
\end{figure}

The black widow system Ter5 O also shows large amplitude X-ray variability as a function of orbital phase Figure~\ref{fig:lcurvebw}, with a $3.5\times10^{-5}$ ($4.0\sigma$) probability of arising from a uniform distribution. Unlike Ter5 P and ad, however, the X-ray minimum occurs around inferior conjunction ($\phi=0.75$). Curiously, no photons are detected during a $\sim$0.1 phase interval\footnote{The non-uniformity in exposure across the binary orbit is at most at the level 16\% so the observed count deficit cannot be explained by incomplete coverage of the orbit.}. 
As noted in Section~\ref{sec:imaging}, the black widow Ter5 ae is not detected as a distinct X-ray source (likely due to source crowding/blending) and the positionally coincident X-ray emission shows no statistically significant evidence of variability at the $P_b=4$\,h binary period.

\begin{deluxetable}{lccrr}
\label{tab:spectra}
\tablecolumns{5} 
\tablewidth{0pt}  
\tabletypesize{\small}
\tablecaption{Summary of spectroscopic analysis of three redback MSPs, \& one black widow MSP, in Terzan 5}  
\tablehead{\colhead{Source} & \colhead{$N_{\rm H}$} & \colhead{$\Gamma$} & \colhead{$L_X$\tablenotemark{a}}  & \colhead{$\chi^2$/dof} \\ 
            \colhead{ }      & \colhead{($10^{22}$\,cm$^{-2}$)}      & \colhead{ } & \colhead{($10^{31}$ erg s$^{-1}$)} &     \colhead{}}
\startdata
Ter5 A\tablenotemark{b}  & ($2.1$)\tablenotemark{c} & $1.24\pm0.69$ & $8.6\pm2.6$ & $0.79/1$ \\
Ter5 P & $2.5^{+0.43}_{-0.38}$ & $0.86^{+0.17}_{-0.16}$ & $32.9^{+2.1}_{-1.8}$ & $74.0/103$ \\
Ter5 O  & $0.8^{+1.0}_{-0.7}$ & $0.73^{+0.73}_{-0.63}$ & $1.7^{+0.35}_{-0.33}$ & $3.55/3$ \\
Ter5ad & $2.2^{+0.6}_{-0.5}$ & $1.16^{+0.24}_{-0.22}$ & $13.4^{+1.6}_{-1.2}$ & $31.2/47$ \\
\enddata
\tablenotetext{a}{Unabsorbed X-ray luminosity in the 0.5--8 keV range in units of $10^{31}$ erg s$^{-1}$ assuming a distance to Terzan 5 of 5.9\,kpc.}
\tablenotetext{b}{Spectrum for Ter5 A of the ``high'' flux state from ObsIDs 13705 and 14478.}
\tablenotetext{c}{Value of $N_{\rm H}$ frozen at nominal cluster value reported by \citep{Bahramian15}.}
\tablenotetext{}{All uncertainties quoted are at a 90\% confidence level.}
\end{deluxetable}

\section{Spectral Properties}
\label{sec:spectral}

As shown in Section~\ref{sec:imaging} and Table~\ref{tab:srclist}, many Terzan 5 MSPs are either marginally detected or have no obvious X-ray counterparts. Nonetheless, these marginal and non-detections still provide useful constraints on the nature of their emission as discussed in Section~\ref{sec:conclusions}. As seen in Figure~\ref{fig:image_xcolor},  these sources have $L_X\lesssim 10^{31}$\,erg\,s$^{-1}$. For the MSPs in the crowded inner core of the cluster, source confusion and chance coincidences are a problem so the measured X-ray properties may not be purely from the MSP. In such an event the estimated luminosities can be considered as upper limits.

Typical MSPs exhibit soft, low-luminosity thermal emission originating from their magnetic polar caps due to heating by a return current from the pulsar magnetosphere. For such source, the emission would contribute $\le$10 counts to the Terzan 5 data set owing to the relatively high hydrogen column density towards the cluster\footnote{An additional factor that hinders the detection of soft sources is the gradual decline in the soft response of ACIS due to contaminant build-up, which strongly impacts the more recent exposures.} ($N_{\rm H}\approx 2\times10^{22}$ cm$^{-2}$) . 

The eclipsing MSPs Ter5 A, O, P, and ad have sufficient counts to permit a formal spectroscopic analysis. For this purpose, the extracted spectra were grouped in photon energy such that each bin contains at least 15 counts. We account for absorption due to the interstellar Galactic medium using the \texttt{tbabs} model in XSPEC, which employs the VERN cross-sections \citep{verner96} and WILM abundances \citep{Wilms00}. Table~\ref{tab:spectra} summarizes the spectroscopy of the redback MSPs Ter5 A, P, and ad, and the black widow Ter5 O.
For Ter5 A, only the data from the ``high'' luminosity state are included and $N_{\rm H}$ is fixed at the nominal cluster value due to the limited number of events in the spectrum. As expected from intra-binary shock emission,  Ter5 P and ad exhibit hard spectra with best fit photon indices $\Gamma\sim 1$. As Ter5 O is considerably fainter, the spectral index is not as well constrained but it does also tend towards a hard spectrum.

\begin{deluxetable*}{lCCCCCl}
\label{tab:redbacks}
\tabletypesize{\small}
\tablecolumns{8} 
\tablewidth{0pt}  
\tablecaption{Summary of X-ray properties of confirmed and candidate redbacks in globular clusters and the field of the Galaxy}  
\tablehead{
\colhead{System} & \colhead{$P$} & \colhead{$P_b$} & \colhead{$D$} & \colhead{Photon} & \colhead{$L_X$\tablenotemark{a}} & \colhead{References\tablenotemark{b}}   \\
\colhead{ }      & \colhead{(ms)} & \colhead{(h)} & \colhead{(kpc)}  & \colhead{index} & \colhead{($10^{31}$ erg\,s$^{-1}$)} &  \colhead{ }    
          }
\startdata
\multicolumn{7}{c}{Globular clusters} \\
\hline
PSR J1748--2446A	 &	11.6	&	1.81	&	5.9	&	1.5\pm2.0	&	<1.3\tablenotemark{c}	&	\citet{Lyne1990}, this work			\\
PSR J1748--2446P	&	1.73	&	8.7	&	5.9	&	0.86\pm0.2	&	32.9\pm2.0	&	\citet{Ransom05}, this work		\\
PSR J1748--2446ad	&	1.40	&	26.3	&	5.9	&	1.16\pm0.2	&	13.4\pm1.4	&	\citet{Hessels06}, this work			\\
Ter5--VLA34	&	?	&	?	&	5.9	&	1.5\pm0.2	&	8.0\pm0.8	&	\citet{Urquhart2020}			\\
Ter5--VLA38	&	?	&	12.3	&	5.9	&	1.3\pm0.1	&	25.5\pm1.3	&	\citet{Urquhart2020}		\\
Ter5--VLA40	&	?	&	?	&	5.9	&	1.3\pm1.0	&	1.5\pm0.7	&	\citet{Urquhart2020}	\\
PSR J0024--7204V (47 Tuc)	&	4.81	&	5.1	&	4.6	&	\nodata	&	\nodata	& \citet{Ridolfi16} \\
PSR J0024--7204W (47 Tuc)	&	2.35	&	3.2	&	4.6	&	1.1\pm0.4	&	4\pm1	& \citet{Freire2001}, \citet{Bogdanov05}			\\
PSR J1740--5340\tablenotemark{e} (NGC 6397)	&	3.65	&	32.5	&	2.4	&	1.6\pm0.2	&	2.2\pm0.3	& \citet{DAmico2001}, \citet{Bogdanov10}		\\
PSR J1824--2452H (M28)	&	4.63	&	10.4	&	5.5	&	1.1\pm0.3	&	1.3\pm1.0	& \citet{Begin_2006}, \citet{Bogdanov2011b}		\\
PSR J1824--2452I\tablenotemark{d} (M28)	&	3.93	&	11	&	5.5	&	1.2\pm0.2	&	22\pm4	& \citet{Papitto13}, \citet{Linares14b}	\\
\hline
\multicolumn{7}{c}{Galactic field} \\
\hline
PSR J1023+0038\tablenotemark{d}	&	1.69	&	4.8	&	1.37	&	1.19\pm0.03	&	9.4\pm0.4	& \citet{Archibald09}, \citet{Bogdanov11}	\\
XSS J12270--4859\tablenotemark{d}	&	1.69	&	6.9	&	1.3	&	1.16\pm0.08	&	9.3\pm0.2	& \citet{Roy2015}, \citet{Bogdanov2014b}			\\
3FGL J0212.1+5320	&	?	&	20.9	&	1.1	&	1.29\pm0.06	&	26.1\pm1.4	& \citet{Linares2017}		\\
1FGL J0523.5--2529	&	?	&	16.5	&	1.1	&	1.10\pm0.8	&	2.9\pm1.5	&  \citet{Strader14}			\\
3FGL J0744.1--2523	&	?	&	2.8	&	1.5	&	\nodata	&	<1.2	& \citet{Salvetti2017}		\\
3FGL J0838.8--2829	&	?	&	5.1	&	1.7	&	1.5\pm0.1	&	6.0\pm0.3	&	\citet{Halpern17}			\\
2FGL J0846.0+2820\tablenotemark{e}	&	?	&	195	&	3.8	    &	(1.1) &	\approx1.8	& \citet{Swihart2017}		\\
3FGL J0954.8--3948	&	?	&	9.3	&	1.7	&	0.8\pm 0.5	&	11\pm5	& \citet{Li2018}		\\
PSR J1048+2339	&	4.88	&	6	&	0.85	&	1.64\pm0.15	&	6.3\pm0.4	& \citet{Deneva2016}, \citet{Yap2019}			\\
PSR J1302--3258	&	3.77	&	18.8	&	1.4	&	\nodata	&	0.38\pm0.09	&	\citet{Hessels2011}		\\
PSR J1306--40	&	2.2	    &	26.3	& 1.2		&	1.31\pm0.04	& 8.4\pm0.2	& \citet{Keane2018}, \citet{Linares2018}		\\
PSR J1417--4402\tablenotemark{e}	&	2.66	&	129	&	1.6	&	1.59\pm0.3	&	26.7\pm4.6	& \citet{Camilo2016}	\\
PSR J1431--4715	&	2.01	&	10.8	&	1.5	&	\nodata	&	\nodata	& \citet{Bates2015}		\\
PSR J1622--0315	&	3.86	&	3.88	&	1.1	&	2.0\pm0.3	&	0.41\pm0.07	& \citet{Gentile2018}			\\
PSR J1628--3205	&	3.21	&	5	    &	1.1	&	1.2\pm0.8	&	2.24\pm0.02	& \citet{Ray2012}, \citet{Roberts2011}			\\
PSR J1723--2837	&	1.86	&	14.8	&	0.9	&	1.13\pm0.02	&	11.4\pm0.4	& \citet{Crawford2013}, \citet{Bogdanov2014}			\\
PSR J1816+4510	&	3.19	&	8.7	    &	4.5	&	(1.5) 	&	1.2\pm0.2	& \citet{Kaplan2012}, \citet{Stovall2014}		\\
PSR J1908+2105	&	2.56	&	3.6	    &	2.6	&	1.3\pm0.5	&	3.5\pm0.7	&	\citet{Gentile2018}			\\
PSR J2039--5617	&	2.65	&	5.5	    &	1.25	&	1.34\pm0.1	&	1.5\pm0.1	& \citet{Clark2020}, \citet{Salvetti2015}		\\
PSR J2129--0429	&	7.61	&	15.2	&	1.4	&	1.10\pm0.1	&	2.0\pm0.1	&	\citet{Hessels2011}, \citet{Al_Noori_2018}			\\
PSR J2215+5135	&	2.61	&	4.2	    &	3.0	&	1.20\pm0.4	&	10.2\pm1.5	&	\citet{Hessels2011}, \citet{Gentile2014}			\\
PSR J2339--0533	&	2.88	&	4.6	&	1.1	&	1.09\pm0.3	&	3.3\pm0.3	& \citet{Romani2011}		\\
\enddata
\tablenotetext{a}{Estimated unabsorbed X-ray luminosity in the 0.5--8 keV band in units of  $10^{31}$ erg\,s$^{-1}$.}
\tablenotetext{b}{First reference reports source discovery; where listed, second reference reports X-ray properties.}
\tablenotetext{c}{Time-averaged luminosity.}
\tablenotetext{d}{Systems known to have undergone transitions between radio pulsar and accretion disk states. The quoted X-ray spectral parameters are for the radio (disk-free) state.}
\tablenotetext{e}{Secondary star is a giant or sub-giant.}
\end{deluxetable*}

\section{Discussion and Conclusions}
\label{sec:conclusions}

We have presented an X-ray imaging, spectroscopy and variability study of the population of rotation-powered MSPs in the Galactic globular cluster Terzan 5, using extensive archival \textit{Chandra} ACIS observations accumulated over 16 years. At least six of these MSPs are coincident with X-ray sources, with more MSPs with plausible but not definitive source matches.  The majority of MSPs either have no X-ray counterparts or have only marginally significant counts above the background, suggesting that these sources have soft X-ray spectra and low luminosities and are rendered either marginally detectable or undetectable due to the high absorbing column towards Terzan 5. For these sources, we can place a conservative $3\sigma$ upper limit on the unabsorbed X-ray luminosity of $\approx 3\times 10^{30}$\,erg\,s$^{-1}$ in the 0.5--8 keV range, assuming a blackbody spectrum with $kT=0.2$\,keV, assuming a distance to Terzan 5 of 5.9\,kpc and $N_{\rm}=2.1\times10^{22}$\,cm$^{-2}$. 

It is interesting to compare the results for Terzan 5 presented here with previous X-ray studies of the globular clusters 47 Tuc \citep{Bogdanov06,Bhattacharya2017} and M28 \citep{Bogdanov11}, which also host sizable populations of known MSPs. In the case of 47 Tuc ($D=4.53$\,kpc), all 25 MSPs are found to have X-ray counterparts in the 540-ks \textit{Chandra} exposure presented by \citet{Bhattacharya2017}, owing to the low intervening column ($3.5\times10^{20}$\,cm$^{-2}$).  The majority of the counterparts exhibit soft, thermal X-ray spectra with temperatures $\sim10^6$\,K and effective radii of order a km for an assumed non-magnetic hydrogen atmosphere. The notable outlier was found to be the redback PSR J0024--7204W, which showed a hard spectrum and orbital variability. A \textit{Chandra} ACIS study of the globular cluster M28 \citep{Bogdanov2011b} based on 237.1\,ks of exposure, resulted in the detection of seven (plus two plausible matches) out of the 12 known M28 pulsars. With the exception of the energetic PSR B1821--24, and the redbacks PSR J1824--2452H and J1824--2452I \citep{Papitto13}, the detected pulsars exhibited relatively soft spectra, with X-ray luminosities $10^{30}-10^{31}$\,\,erg\,s$^{-1}$(0.3--8 keV). Thus, in terms of X-ray properties, the Terzan 5 sample appears to be representative of cluster MSPs. Moreover, these X-ray faint globular cluster MSPs share similarities with nearly all nearby MSPs ($\lesssim$1\,kpc) observed in the field of the Galaxy, which tend to show soft, thermal pulsations \citep{Zavlin06,Ray19,Guillot2019}, with X-ray luminosities reaching down to $L_X=10^{29}$\,\,erg\,s$^{-1}$ \citep{Swiggum2017}.

The X-ray counterparts of the redbacks Ter 5 A, P and ad, and the black widow Ter 5 O, stand out from the Terzan 5 population in terms of X-ray spectra and variability, which possess the hallmarks of intrabinary shock radiation. It is generally understood that the observed non-thermal X-rays from redback (and some black widow) binaries originates in a synchrotron-emitting standing shock driven by the interaction of the relativistic pulsar wind with outflowing material from the close companion star \citep{Arons93}. The observed X-ray radiation typically exhibits a hard non-thermal spectrum with power-law photon index of $\Gamma\approx1$ \citep{Bogdanov11,Bogdanov2014} and unabsorbed X-ray luminosities in the range of $10^{31-32}$\,erg\,s$^{-1}$ (averaged over the orbital cycle). For systems where good orbital coverage is available in X-rays, a universal characteristic is a factor of $2-3$ variation in X-ray luminosity, with an X-ray minimum at superior conjunction (when the companion is between the pulsar and observer) and a pronounced single or double horned maximum (as seen in Figure~\ref{fig:lcurves}).This orbital variability  can be interpreted as being due to the the changing view of the shock, with the sharp features near the flux maximum being caused by relativistic beaming when the observer line of sight is tangent to the shock surface. 

The observed shock emission can, in principle, serve as a powerful diagnostic of rotation-powered MSP winds and collisionless shocks. \citet{Romani16} and \citet{Wadiasingh17} have undertaken efforts to develop sophisticated semi-analytic models to gain insight from the phenomenology of this wind interaction. These models qualitatively capture the broad features of the multi-wavelength data, in particular the X-ray light curves and the radio eclipse durations, although there appear to be additional unmodeled physical effects that require further investigation. The X-rays/$\gamma$-rays generated by the shock alone do not supply sufficient power to account for the heating of the face of the secondary star, and necessitates a mechanism that channels the post-shock flow to the surface of the companion like magnetic ``ducts'' that connect to the stellar surface \citep{Sanchez2017}. Another peculiar finding of these studies is that reproducing the observed X-ray light curves of redbacks, with an X-ray minimum at $\phi\approx0.25$, requires a wind shock interface that is closer to and sweeps around the pulsar. The shock geometry and sweep orientation are determined primarily by the momentum flux ratio of the two stellar outflows and the ratio between the massive companion wind speed and the orbital velocity. Thus, for redbacks, including Ter 5 P and ad, in this interpretation the momentum flux of the companion wind appears to be dominant.

In contrast, the black widow Ter 5 O has an X-ray minimum at $\phi\approx0.75$, the same as the original black widow, PSR B1957+20 \citep{Huang2012}. Based on the models of \citet{Romani16} and \citet{Wadiasingh17}, this can be interpreted as a shock sweeping around the secondary star, presumably due to much weaker outflows from the lower mass secondary stars in these systems.  The sample of black widow systems with detailed X-ray orbital light curves is currently very limited \citep[see, e.g.,][]{Gentile2014,Arumugasamy2015}, so at present we cannot establish whether this is a ubiquitous property of this class of eclipsing MSPs.

The number of redback MSPs and strong candidates has grown dramatically in recent years, in large part due to targeted radio, optical, and X-ray searches of unassociated \textit{Fermi} Large Area Telescope $\gamma$-ray sources \citep[e.g.,][]{Ray2012,Strader14}. In Table~\ref{tab:redbacks} we have compiled the spin, orbital and X-ray properties (where available) of the current sample of published redbacks in globular clusters and the field of the Galaxy. For a number of targets, there is no detection of MSP pulsations, but they exhibit multi-wavelength characteristics that closely match those of redbacks with confirmed MSP primaries.  Ter5 A, P, and ad, when taken together with the tMSP candidate CX1 \citep{Bahramian2018} and three radio-selected redback candidates identified by \citet{Urquhart2020},  account for a significant fraction of systems of this variety. Ter5 P, in particular, stands out as the most X-ray luminous redback known, with Ter5 ad and Ter5-VLA38 following close behind along with the three confirmed tMSPs, with $L_X>1\times10^{32}$ erg\,s$^{-1}$. Also of note is that most X-ray luminous redbacks tend to have spin periods $\lesssim 2$\,ms, which is likely due to high spin-down luminosities\footnote{The determination of the intrinsic spin-down rate $\dot{P}$ and hence $\dot{E}$ of globular cluster MSPs is difficult because the pulsars experience acceleration due to the cluster potential, which can dominate the measured value \citep[see, e.g.,][]{Freire2001}.} $\dot{E}$ since $\dot{E}\propto P^{-3}$. However, the observed X-ray properties such as the shock luminosity, broad band spectrum, and orbital light curve morphology likely depend sensitively on a combination of additional factors such as the orientation of the pulsar wind outflow relative to the orbital plane, the companion mass, its particle outflow and magnetic field strength/structure, the orbital separation, and viewing geometry.  Thus, further detailed modeling of the X-ray (and optical, where available) light curves of redbacks as an ensemble is require to gain further insight regarding the various factors that determine the apparent characteristics of the intra-binary shock.    
The X-ray flux behavior of Ter5 A is reminiscent of that observed from PSR J1048$+$2339, a confirmed redback in the field of the Galaxy, and XMMU J083850.38$-$282756.8, a putative redback binary and the likely optical and X-ray counterpart of the \textit{Fermi} LAT source 3FGL J0838.8$-$2829 \citep{Halpern17,Cho18}. These systems have been observed to undergo short-lived X-ray and optical flaring states, which can be attributed to episodic strong magnetic activity on the face of the companion heated by the pulsar wind and/or shock. In principle, the intermittent spikes in X-ray luminosity observed for Ter5 A could be due to the presence of an accretion disk. However, the variability pattern does not resemble the moding behavior seen in tMSPs, although the tMSP sample is very limited so we may not have seen the full range of possible behavior from such systems. In any event, this MSP along with the rest of the sizable redback population of Terzan 5 warrant further monitoring as there is a high likelihood of catching a state transition from at least one of them.

\acknowledgments
Support for this project was provided by the NASA through Chandra Award Number AR6-17004X issued by the Chandra X-ray Center, which is operated by the Smithsonian Astrophysical Observatory for and on behalf of NASA under contract NAS8-03060.  COH is supported in part by NSERC Discovery Grant RGPIN-2016-04602. This work has made use of data obtained from the Chandra Data Archive and software provided by the Chandra X-ray Center (CXC) in the application package CIAO. This research has made use of NASA's Astrophysics Data System (ADS) Bibliographic Services and the arXiv.

\facilities{\textit{Chandra} ACIS} 
\software{CIAO \citep{2006SPIE.6270E..1VF}, ACIS\_EXTRACT \citep{Broos2010}, SciPy  \citep{2020SciPy-NMeth}, NumPy \citep{2011CSE....13b..22V}, Matplotlib \citep{4160265}}

\bibliographystyle{yahapj}
\bibliography{references}

\end{document}